\documentclass{MYws-ijmpa}
\usepackage[super,compress]{cite}
\usepackage{graphicx}
\usepackage{url,hyperref}
\begin{document}
\markboth{Antonio Pich}{Tau Physics Opportunities at the STCF}

%
\catchline{}{}{}{}{}
%

\title{Tau Physics Opportunities at the Super Tau-Charm Facility} 

\author{Antonio Pich}

\address{IFIC, Universitat de València – CSIC, Catedrático José Beltrán 2, E-46980 Paterna, Spain\\ Antonio.Pich@ific.uv.es}

\maketitle


\begin{abstract}
The super tau-charm facility will provide excellent conditions to perform a high-precision investigation of the tau-lepton properties: very high statistics, controllable systematics and low backgrounds. An overview of the broad physics program that could be addressed at this facility is presented.

\keywords{Tau properties; electroweak interactions; QCD.}
\end{abstract}

\ccode{PACS numbers: 14.60.Fg  13.35.-r  12.15.-y  12.38.-t}


\section{Advantages of the threshold region}

High-precision measurements of the $\tau$ properties allow us to perform stringent tests of the Standard Model (SM) and search for possible signals of new phenomena\cite{Pich:2013lsa}. 
Being the heaviest known lepton, the $\tau$ is expected to be the most sensitive one to the unknown flavour dynamics responsible for the fermion replication. Moreover, it is the only lepton able to decay into hadrons, which makes possible to study the low-energy regime of QCD in rather clean (semileptonic) conditions. 

$\tau^+\tau^-$ pairs are copiously produced in any electron-positron collider running above the $\sqrt{s_{\mathrm{th}}}= 2m_\tau$ threshold. Therefore, the physics of the $\tau$ lepton can be investigated in a broad variety of energy regimes. Each energy setting corresponds to
different experimental conditions with its own advantages and disadvantages.
Table~\ref{tab:TauProd} compares the expected number of $\tau^+\tau^-$ events that will be collected at Belle II,\cite{Belle-II:2018jsg} the super tau-charm facility (STCF)\cite{Achasov:2023gey} and a future Tera-Z facility\cite{FCC:2018byv} with the total data samples already accumulated by the ALEPH experiment at LEP,\cite{ALEPH:2005qgp} and by BaBar and Belle at the B factories.\cite{BaBar:2014omp}

\begin{table}[tbh]
\tbl{$\tau$ data samples collected at ALEPH\cite{ALEPH:2005qgp} and the B factories,\cite{BaBar:2014omp} compared with the expected number of events at Belle II,\cite{Belle-II:2018jsg} the STCF\cite{Achasov:2023gey} and a  future 
Tera-Z facility.\cite{FCC:2018byv}}
{\begin{tabular}{@{}ccc@{}} \toprule
Experiment & Energy setting & \# of events \\  \colrule
ALEPH & $M_Z$ & $3.3\cdot 10^5$ reconstructed $\tau$ decays \\
BaBar / Belle & $\Upsilon$ & $1.4\cdot 10^9$ $\tau^+\tau^-$ pairs \\
Belle II & $\Upsilon$ & $4.6\cdot 10^{10}$ $\tau^+\tau^-$ pairs \\
STCF & $\Psi$ & $2.1\cdot 10^{10}$ $\tau^+\tau^-$ pairs ($10^8$ near threshold)\\
Tera-Z & $M_Z$ & $1.7\cdot 10^{11}$ $\tau^+\tau^-$ pairs 
\\ \botrule
\end{tabular} \label{tab:TauProd}}
\end{table}

The $\tau^+\tau^-$ data samples collected at LEP are much smaller than the ones available from the B factories. For instance, the ALEPH experiment only accumulated $3.3\cdot 10^5$ reconstructed $\tau$ decays, to be compared with $1.4\cdot 10^9$ $\tau^+\tau^-$ pairs recorded by BaBar and Belle. In spite of this fact, the experimental values of many $\tau$ properties, such as the main branching ratios or the inclusive spectral distributions, are still dominated by the LEP data.
The special kinematic configuration of the back-to-back boosted $\tau^+\tau^-$ pairs emerging from a Z decaying at rest provides a friendly scenario for $\tau$ analyses with quite simple tagging procedures. The Tera-Z option of a future FCC-ee collider running at the $Z$ peak would certainly benefit from an enormous data sample, in extremely clean kinematic conditions.\cite{Pich:2020qna}

At the B factories there are large backgrounds from bottom and charm decays, but the huge statistics that will be collected by the on-going Belle II experiment will make possible a much tighter control of the experimental systematics, sharpening the measurement strategies and opening a broad range of interesting opportunities for precision $\tau$ physics.\cite{Bodrov:2024wrw}
 A comparable data sample will be available at the STCF, offering a nice complementarity with different backgrounds (no $b$ quarks) and symmetric beam energies. Even more important is the ability of the STCF to run between the $\tau^+\tau^-$ and $D^0\bar D^0$ production thresholds, an energy range that is free from heavy-quark backgrounds. Moreover, the light-quark backgrounds can be experimentally measured below the $\tau^+\tau^-$ threshold, while the narrow $J/\psi$ and $\psi(2S)$ peaks can be used for calibration. At the production threshold,  more than $10^8$ $\tau^+\tau^-$ events per year could be collected, making possible a very significant breakthrough in precision physics. Measurements less sensitive to backgrounds and systematics will profit from the larger statistics available at higher energies (the highest $\tau^+\tau^-$ production cross section is reached around $\sqrt{s}\approx 4.2$~GeV).

Running at threshold provides the most efficient way to measure precisely the $\tau$ mass, as demonstrated long time ago by the spectacular 1992 BES determination of $m_\tau$, with only seven identified events.\cite{BES:1992xnu} With monochromatized beams, an improved precision of $\Delta m_\tau\sim 0.02$~MeV is achievable at the STCF.\cite{Achasov:2023gey} This could also make possible the experimental observation of ortho-ditauonium ($\mathcal{T}_1$), a spin-1 $\tau^+\tau^-$ bound state, via $e^+e^-\to \mathcal{T}_1\to\mu^+\mu^-$.\cite{dEnterria:2023yao}. The threshold region offers also an improved sensitivity to two-body decays, such as $\tau^-\to\nu_\tau\pi^-$, $\tau^-\to \nu_\tau K^-$ or $\tau^-\to\ell^-\gamma$ ($\ell = e,\mu$), due to the monochromatic energies of the final particles.

\section{Leptonic Decays}

\begin{figure}[tb]
\centerline{\includegraphics[width=8cm]{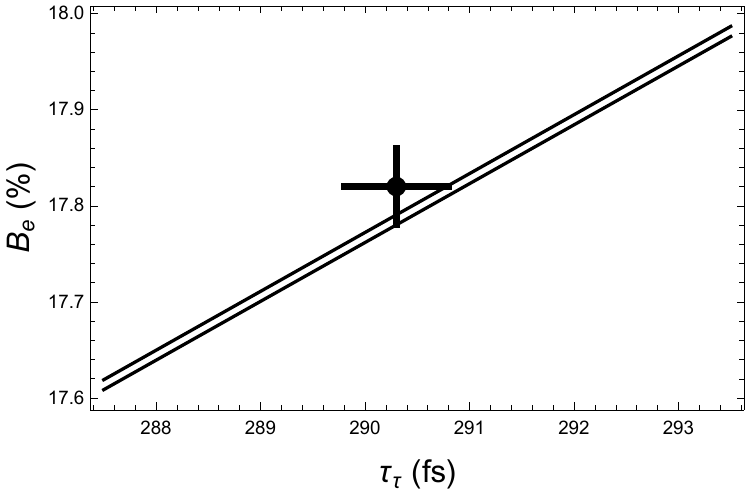}}
\caption{SM relation between $B_e$ and $\tau_\tau$ (diagonal band), compared with their experimental values. \label{fig:BTplot}}
\end{figure}

In the SM, the $\tau$ lepton decays through the emission of a virtual $W$ boson, i.e., $\tau^-\to\nu_\tau W^{-*}\to\nu_\tau X^-$ with $X^- = e^-\bar\nu_e, \mu^-\bar\nu_\mu, d\bar u, s \bar u$. The universal strength of the charged-current interaction is precisely determined from the $\mu^-\to e^-\bar\nu_e\nu_\mu$ decay, which results in an accurate prediction for the leptonic decay widths of the $\tau$. As shown in 
Figure~\ref{fig:BTplot}, this implies a relation between the $\tau$ lifetime and its electronic branching ratio, $B_e =\tau_\tau \,\Gamma_{\tau\to e}$, that
is in excellent agreement with the current experimental values. The width of the SM band in the figure reflects the present uncertainty on the $\tau$ mass ($\Gamma_{\tau\to e}\propto m_\tau^5$).\cite{ParticleDataGroup:2022pth,Belle-II:2023izd}
More precise measurements of $B_e$ and $\tau_\tau$ at the STCF and Belle II will strengthen the significance of this SM test.

The ratio between the $\tau\to\mu$ and $\tau\to e$ leptonic widths is accurately predicted by the SM to be $B_\mu/B_e = 0.972563\; (3)$, which exhibits a slight tension ($1.3\,\sigma$) with the experimental value quoted by the PDG: $(B_\mu/B_e)_{\mathrm{PDG}} = 0.9762\; (28)$ \cite{ParticleDataGroup:2022pth}. The small difference originates in a 2010 BaBar measurement of this ratio 
that is $1.8\,\sigma$ higher than the SM expectation.\cite{BaBar:2009lyd} 
Belle II has recently made public a measurement of this ratio (the most precise up to now),\cite{Belle-II:2024vvr}
$(B_\mu/B_e)_{\mathrm{Belle\, II}} = 0.9675\; (37)$, leading to a new world average value
$(B_\mu/B_e) = 0.9730\; (22)$, in perfect agreement with the SM prediction.


\subsection{Lepton universality}

Comparing the measured leptonic decay widths of the $\tau$ and the $\mu$, one can test the flavour universality of the $W$ couplings, i.e., that $g_e=g_\mu=g_\tau\equiv g$. Complementary tests can be obtained from the hadronic decay modes $\tau\to\nu_\tau\pi$ and $\tau\to\nu_\tau K$, 
from leptonic (and semileptonic) pion and kaon decays, and from the direct leptonic decays of the $W$ boson. The most accurate experimental determinations of the ratios $|g_\ell/g_{\ell'}|$ are shown in Table~\ref{tab:universality} (tests from heavy-meson decays are less precise). The leptonic $\tau$ decays confirm the universality of the three lepton couplings at the 0.14\% level, while a better 0.09\% accuracy on $|g_\mu/g_e|$ is obtained from the ratio of the two leptonic $\pi$ decay modes. 

\begin{table}[htb]\centering
\tbl{Experimental determinations of the ratios \ $|g_\ell/g_{\ell'}|$.\cite{Pich:2013lsa,ParticleDataGroup:2022pth,Belle-II:2024vvr,Arroyo-Urena:2021nil}}
{\begin{tabular}{@{}cccccc@{}} \toprule  
&
 $\Gamma_{\tau\to e}/\Gamma_{\mu\to e}$ &
 $\Gamma_{\tau\to\pi}/\Gamma_{\pi\to\mu}$ &
 $\Gamma_{\tau\to K}/\Gamma_{K\to\mu}$ &
 $\Gamma_{W\to\tau}/\Gamma_{W\to\mu}$
\\ 
 $|g_\tau/g_\mu|$
 & $1.0009\; (14)$ & $0.9964\; (38)$ & $0.986\; (8)$ & $1.001\; (10)$
\\ \colrule
&
 $\Gamma_{\tau\to\mu}/\Gamma_{\tau\to e}$ &
 $\Gamma_{\pi\to\mu} /\Gamma_{\pi\to e}$ &
 $\Gamma_{K\to\mu} /\Gamma_{K\to e}$ &
 $\Gamma_{K\to\pi\mu} /\Gamma_{K\to\pi e}$ 
\\ 
 $|g_\mu/g_e|$
 & $1.0002\; (11)$ & $1.0010\; (9)$ & $0.9978\; (18)$ & $1.0010\; (25)$ 
\\ 
\\[-7pt]\hline &  \multicolumn{1}{c|}{} &&&\\[-6pt]
& \multicolumn{1}{c|}{$\Gamma_{W\to\mu} /\Gamma_{W\to e}$} &&
 $\Gamma_{\tau\to\mu}/\Gamma_{\mu\to e}$
 & $\Gamma_{W\to\tau}/\Gamma_{W\to e}$
\\ 
 $|g_\mu/g_e|$ & \multicolumn{1}{c|}{$1.001\; (3)$} &
 $|g_\tau/g_e|$
 & $1.0027\; (14)$ & $1.007\; (10)$
\\[-7pt] &  \multicolumn{1}{c|}{} &&&\\ \Hline
\end{tabular}\label{tab:universality}}
\end{table}

New physics beyond the SM (BSM) would contribute differently to each decay mode. Therefore, the different tests in the table provide complementary and very valuable information. The experimental uncertainties on $|g_\tau/g_\mu|$ from $\Gamma_{\tau\to\pi}/\Gamma_{\pi\to\mu}$ and  $\Gamma_{\tau\to K}/\Gamma_{K\to\mu}$ are completely dominated by the present uncertainties on the $\tau\to\pi$ ($0.5\%$) and $\tau\to K$ ($1.4\%$) branching fractions, while theoretical errors on these hadronic ratios are currently at the $0.28\%$ level.\cite{Arroyo-Urena:2021nil} More accurate $\tau$-decay measurements and a better theoretical control of radiative corrections are needed to significantly improve the precision of these hadronic tests.

The LEP data exhibited a $2.7\,\sigma$ excess in $W\to\tau\nu$ decays. Thanks to recent and more precise LHC measurements, the direct information from $W$ decays is now also in agreement with universality, although the reached accuracy is only 1\% (0.3\%) for $|g_\tau/g_{\mu,e}|$ ($|g_\mu/g_e|$).

\subsection{Lorentz structure}

The effective (dimension-6) four-lepton interaction Hamiltonian
\begin{equation}
\label{eq:hamiltonian}
{\cal H} \; =\;  4\, \frac{G_{\ell'\ell}}{\sqrt{2}}\;
\sum_{n,\epsilon,\omega}\,
g^n_{\epsilon\omega}\,
\left[ \overline{\ell'_\epsilon}
\Gamma^n {(\nu_{\ell'})}_\sigma \right]\,
\left[ \overline{({\nu_\ell})_\lambda} \Gamma_n
	\ell_\omega \right]
\end{equation}
parametrizes the most general (lepton-number conserving) $\ell^-\to\nu_\ell\ell'^-\bar\nu_{\ell'}$ decay amplitude consistent with locality and Lorentz invariance.\cite{Pich:2013lsa}
There are ten possible operators with their corresponding complex couplings $g^n_{\epsilon\omega}$, where $n=S,V,T$ indicates the type of interaction (scalar, vector, tensor), while the subindices $\omega$ and $\epsilon$ label the chiralities (left or right) of 
$\ell$ and $\ell'$, respectively. Taking out the global normalization factor $G_{\ell'\ell}$ that is determined by the total decay rate, these couplings are bounded to the ranges $|g^S_{\epsilon\omega}|\le 2$, $|g^V_{\epsilon\omega}|\le 1$ and $|g^T_{\epsilon\omega}|\le 1/\sqrt{3}$. In the SM, $g^V_{LL}=1$, while all other couplings are identically zero. Any additional contribution from new physics at higher scales would result in non-zero values for some of these effective couplings. 

Measuring the energy and angular distribution of the final charged lepton, complemented with polarisation information whenever available, one can disentangle the different contributions
(information from the inverse transition $\nu_\ell \ell'^-\to\ell^-\nu_{\ell'}$ is also needed).
This has been successfully achieved in $\mu$ decay, demonstrating that the bulk of the decay amplitude is indeed of the predicted $V-A$ type, $|g^V_{LL}| > 0.960$ (90\% C.L.), and establishing tight upper bounds on all other couplings.\cite{ParticleDataGroup:2022pth}

A model-independent analysis is more challenging for the $\tau$ because of its much shorter lifetime. The current constraints on the $\tau\to\mu$ couplings are shown in Table~\ref{tab:Michel_tau} (the bounds for $\tau\to e$ are very similar). The correlated distribution of the $\tau^+\tau^-$ pairs produced in $e^+e^-$ annihilation is sensitive to the $\tau$ polarisation. Taking advantage of the polarization provided by the $Z$ decay, the LEP experiments 
were able to put upper limits on those couplings with an initial right-handed $\tau$. However, the Lorentz structure of a left-handed decaying $\tau$ remains undetermined.

\begin{table}[tbh]\centering
\tbl{Experimental limits on the $\tau^-\to \mu^-\bar\nu_\mu\nu_\tau$ couplings ($95\%$ C.L.).\cite{ParticleDataGroup:2022pth}\hfill\mbox{}}
{\begin{tabular}{@{}cccc@{}} \toprule  
 $|g_{RR}^S| < 0.72$ & $|g_{LR}^S| < 0.95$ & $|g_{RL}^S| \leq 2$
 & $|g_{LL}^S| \leq 2$ \\[2pt]
 $|g_{RR}^V| < 0.18$ & $|g_{LR}^V| < 0.12$ & $|g_{RL}^V| < 0.52$
 & $|g_{LL}^V| \leq 1$ \\[2pt]
 $|g_{RR}^T| \equiv 0$ & $|g_{LR}^T| < 0.079$ & $|g_{RL}^T| < 0.51$
 & $|g_{LL}^T|\equiv 0$ 
 \\ \botrule
 \end{tabular}}\label{tab:Michel_tau}
\end{table}

Quite recently, the Belle collaboration has accomplished the first polarization measurement of the secondary charged lepton in the $\tau\to\mu$ decay. The huge available statistics ($\sim 9\cdot 10^8$ $\tau^+\tau^-$ pairs) compensates the very small probability to detect a muon decaying inside the detector so that it has become possible to collect 165 signal-candidate events with a reconstructed secondary decay. From the angular distribution of the final electrons one can then measure the Michel parameter $\xi'= 0.22\pm 0.94\pm 0.42$,\cite{Belle:2023udc}
which determines the longitudinal polarization of the daughter $\mu$ for an unpolarized $\tau$:
\begin{equation}
\frac{1}{2}\left(1-\xi'\right) = \frac{1}{4}\left(|g^S_{RR}|^2 + |g^S_{RL}|^2\right) + 3\, |g^T_{RL}|^2 + |g^V_{RR}|^2 + |g^S_{RL}|^2
\equiv\mathcal{Q}_{RR} + \mathcal{Q}_{RL} \equiv \mathcal{Q}_{\mu_R}\, .
\end{equation}
The total decay width is proportional to $\mathcal{Q}_{\mu_R}+\mathcal{Q}_{\mu_L}\equiv 1$, which has been normalized to one so that $\mathcal{Q}_{\mu_R}$ and $\mathcal{Q}_{\mu_L}$ can be interpreted as the total probabilities to decay into a right-handed or left-handed muon, respectively.
Since $0\le\mathcal{Q}_{\mu_R}\le 1$, the total range of variation of $\xi'$ is $[-1,+1]$ ($\xi'=1$ in the SM). The Belle measurement of $\xi'$ implies $\mathcal{Q}_{\mu_R} < 1.23$ (90\% C.L.), which unfortunately does not yet constrain the effective Hamiltonian. The experimental error is dominated by statistics; therefore, both Belle II and the STCF could provide competitive and very useful determinations of $\xi'$. Notice that an upper limit on $(1-\xi')$ implies upper bounds on five different effective couplings.

A first study of the precision achievable at the STCF for the $\tau\to\mu$ Michel parameters sensitive to the polarization of the daughter muon
has been presented at this workshop.\cite{Pakhlov:STCF}. Assuming 80\% positron polarization and that the more common
Michel parameters $\rho$, $\eta$, $\xi$ and $\xi\delta$ are measured with $10^{-3}$ accuracy from the energy distribution, one gets the estimated sensitivities shown in Table~\ref{tab:MichelSensitivities}, which compare favourably with those obtained in muon decay.

\begin{table}[tbh]\centering
\tbl{Estimated sensitivities to the $\mu$-polarization Michel parameters of $\tau\to\mu$ at the STCF, compared with the current values in $\mu$ decay.\cite{Pakhlov:STCF}\hfill\mbox{}}
{\begin{tabular}{@{}cccc@{}} \toprule  
 Parameter & SM value & $\mu^-\to e^-\bar\nu_e\nu_\mu$ & $\tau^-\to \mu^-\bar\nu_\mu\nu_\tau$ 
\\  \colrule 
$\xi'$ & 1 & $1.00\pm 0.04$ & $?\pm 0.006$
\\
$\xi''$ & 1 & $0.98\pm 0.04$ & $?\pm 0.03$
\\
$\alpha'/A$ & 0 & $-0.010\pm 0.020$ & $?\pm 0.014$
\\
$\beta'/A$ & 0 & $0.002\pm 0.007$ & $?\pm 0.007$
\\ \botrule
\end{tabular}}\label{tab:MichelSensitivities}
\end{table}

\section{Hadronic Decays}

The $\tau$ lepton has a large variety of kinematically-allowed semileptonic decay modes, $\tau^-\to\nu_\tau H^-$, which probe the matrix elements of the vector and axial-vector quark currents between the vacuum and the final hadronic states $H^-$. Identifying the strangeness of the produced hadrons, one can also disentangle the Cabibbo allowed and suppressed quark currents. Thus, the $\tau$ lepton provides an excellent data sample to investigate the low-energy dynamics of QCD in the resonance region around 1~GeV.

The hadronic matrix elements governing the lowest-multiplicity decays $\tau^-\to\nu_\tau\pi^-$ and $\tau^-\to\nu_\tau K^-$ are already known from $\pi^-\to\mu^-\bar\nu_\mu$ and $K^-\to\mu^-\bar\nu_\mu$. Taking appropriate ratios of the corresponding decay amplitudes, one gets then the universality tests shown in Table~\ref{tab:universality}. Using the lattice calculation of the kaon decay constant, one can also extract $|V_{us}|$ from the $\tau^-\to\nu_\tau K^-$ decay width, but the current result is not yet competitive with the determinations from $K_{\ell 2}$ and $K_{\ell 3}$ because of the much larger experimental uncertainty on the $\tau\to K$ branching ratio.

Hadronic decays into two or more pseudoscalars ($\pi$, $K$, $\eta$) are sensitive to a broad spectrum of form factors that carry interesting dynamical information. The high statistics that will be collected at the STCF will make possible a very detailed study of differential distributions and
a precise measurement of high-multiplicity and Cabibbo-suppressed decays. Highly-suppressed decays such as the odd G-parity final state $\pi^-\eta$ could also be searched for; an observation of this decay mode above its expected branching ratio of around $10^{-5}$ would imply new physics incorporating second-class currents.\cite{Pich:1987qq,Escribano:2016ntp,Moussallam:2021flg}

\subsection{Spectral functions}

The inclusive invariant-mass distributions of the final hadrons in $\tau$ decay measure the absorptive parts (spectral functions) of the two-point correlation functions of the vector and axial-vector quark currents, which provide direct information about the fundamental QCD dynamics. An accurate determination of these hadronic distributions requires a tight control of experimental systematic uncertainties without introducing relative biases (normalizations, acceptances) among the different final states. The favourable background conditions of the STCF will make possible to perform a significant improvement of the currently available spectral functions, which were extracted from the old and statistically-limited LEP data.

Using short-distance QCD techniques, the weighted integrals of these spectral functions can be theoretically predicted with high accuracy. The perturbative contributions are
already known with an impressive four-loop accuracy,\cite{Baikov:2008jh} i.e. at $\mathcal{O}(\alpha_s^4)$,
while non-perturbative corrections are suppressed by at least four powers of the $\tau$ mass (the exact power depends on the weight function adopted).\cite{Pich:2020gzz}

The Cabibbo-allowed component of the $\tau$ hadronic width turns out to be very sensitive to the strong coupling.\cite{Braaten:1991qm,Narison:1988ni} The perturbative QCD contribution enhances the naive parton model expectation by a sizeable 20\%, while non-perturbative corrections are heavily suppressed by a factor $m_\tau^6$. Moreover, from the measured hadronic distribution it is possible to fix the small non-perturbative corrections, which results in a precise determination of the QCD coupling at the $\tau$ mass scale. Recent analyses of the updated ALEPH spectral functions give\cite{Davier:2013sfa,Pich:2016bdg}
\begin{equation}
\alpha_s^{(n_f=3)}(m_\tau^2) = 0.328\pm 0.013
\qquad\longrightarrow\qquad
\alpha_s^{(n_f=5)}(M_Z^2) = 0.1197\pm 0.0015\, .
\end{equation}
The beautiful agreement with the value extracted from the $Z$ hadronic width,
$\alpha_s^{(n_f=5)}(M_Z^2) = 0.1199\pm 0.0029$, confirms in a very significant way the predicted energy dependence of the running QCD coupling.

Improved measurements of the spectral functions would bring a better understanding of the infra-red QCD dynamics, allowing us to pin down more precisely the non-perturbative corrections. Larger data samples are needed to access the upper end of the kinematically-allowed energies, which is currently hampering the precision of theoretical analyses because of the sizeable experimental uncertainties.

The hadronic spectral distributions contain also precious information on the non-trivial structure of the QCD vacuum. In the massless quark limit, the vector and axial-vector correlation functions are identical to all orders in perturbation theory, reflecting the underlying chiral symmetry of the QCD Lagrangian. Therefore, the weighted integrals of the difference between the vector and axial spectral functions are pure non-perturbative observables, which only receive non-zero contributions from dynamical operators that break chiral symmetry. They can be used to determine important parameters characterizing the QCD vacuum and several low-energy couplings of Chiral Perturbation Theory, the low-energy effective theory of the QCD Golsdstone bosons.\cite{Pich:2020gzz,Pich:2021yll,Gonzalez-Alonso:2016ndl,Boito:2015fra}

\subsection{Muon anomalous magnetic moment}

The SM prediction for $a_\mu^{\mathrm{hvp}}$, the hadronic-vacuum-polarization contribution to the muon $g-2$, is obtained from a weighted integral of the electromagnetic (vector) spectral function, which can be extracted from the available data on $e^+ e^-\to\mathrm{hadrons}$.\cite{Aoyama:2020ynm} Since the particular weight appearing in this integral emphasizes the very low-energy region, the numerical result is dominated ($75\%$) by the $2\pi$ final state. 

For $s\le m_\tau^2$, the isospin-1 contribution to $a_\mu^{\mathrm{hvp}}$ can also be obtained from $\tau$ decay data because the charged and neutral vector currents are related by an isospin rotation. Taking into account the small isospin-breaking corrections,\cite{Cirigliano:2001er,Cirigliano:2002pv} and complementing with $e^+e^-$ data the missing contributions from higher energies, the $\tau$ data imply a higher value of $a_\mu^{\mathrm{hvp}}$ than the one obtained from $e^+e^-$.\cite{Davier:2010fmf,Davier:2010nc,Miranda:2020wdg} In fact, from the $e^+e^-\to\pi^+\pi^-$ data one predicts a $\tau^-\to\nu_\tau\pi^-\pi^0$ branching ratio smaller than its measured experimental value,\cite{Davier:2010fmf} which suggests some unaccounted systematic effect in the quoted uncertainties.

The most recent lattice evaluations \cite{Borsanyi:2020mff,Ce:2022kxy,Kuberski:2024bcj,ExtendedTwistedMass:2022jpw,RBC:2023pvn} fully support a higher value of $a_\mu^{\mathrm{hvp}}$, as suggested long time ago by the $\tau$ data and in better agreement with the experimental determination of the muon $g-2$.\cite{Muong-2:2023cdq} This has been further corroborated by a recent analysis of the Euclidean Adler function (the logarithmic derivative of the electromagnetic two-point function) with $e^+e^-$, $\tau$ and lattice data, which shows that the $e^+e^-$ results are systematically lower than the QCD predictions while the $\tau$ and lattice determinations are in nice agreement with QCD.\cite{Davier:2023hhn} 

The very recent high-precision measurement of  $e^+e^-\to\pi^+\pi^-$ by the CMD-3 collaboration,\cite{CMD-3:2023alj,CMD-3:2023rfe} finds a pion form factor larger than previous $e^+e^-$ averages in the whole energy range measured, confirming the previous indications and showing the need for new high-precision measurements of the spectral functions both in $e^+e^-$ annihilation and $\tau$ decay. Obviously, the STCF could make a very significant breakthrough in this enterprise, which is needed to improve the SM predictions of $(g-2)_\mu$ and $\alpha_{\mathrm{QED}}(M_Z)$.

\subsection{$V_{us}$}   

The ratio of the inclusive $\tau$ decay widths into the Cabibbo-suppressed ($\Delta S=1$) and Cabibbo-allowed ($\Delta S=0$) final states provides a clean determination of $V_{us}$, which does not involve theoretically-estimated form factors or decay constants.\cite{Gamiz:2004ar,Gamiz:2002nu} In the  limit of exact SU(3) symmetry, this ratio directly fixes $|V_{us}/V_{ud}|$. Taking into account the small
SU(3)-breaking correction\cite{Gamiz:2006xx,Pich:1999hc,Pich:1998yn} $\delta R_{\tau,\mathrm{th}} = 0.240 \pm 0.032$ and the PDG value of $V_{ud}$,\cite{ParticleDataGroup:2022pth} the current $\tau$ data imply\cite{HFLAV:2022esi}
\begin{equation}
|V_{us}| = \left( \frac{R_{\tau,S}}{\frac{R_{\tau,V+A}}{|V_{ud}|^2}- \delta R_{\tau,\mathrm{th}}}
\right)^{1/2} = 0.2184\pm 0.0021\, ,
\end{equation}
where $R_{\tau,S}$ and $R_{\tau,V+A}$ are the Cabibbo suppressed and allowed $\tau$ decay widths
normalized to the electronic width. This result is lower by $3.8\,\sigma$ than the CKM-unitarity 
expectation $|V_{us}|^{\mathrm{uni}} = \sqrt{1-|V_{ud}|^2-|V_{ub}|^2} = 0.2277\pm 0.0013$. This seems to reinforce the so-called Cabibbo anomaly that arises when combining the kaon determinations of $|V_{us}|$ with the  value of $|V_{ud}|$ from nuclear $\beta$ decays.\cite{Passemar:STCF,Kitahara:STCF}

However, one should keep in mind that the only fully-inclusive measurement of $R_{\tau,S}$ was done at LEP with a very scarce statistics. Moreover, the exclusive branching rations measured by BaBar and Belle are on average slightly smaller than the LEP ones, a systematic effect that is not yet well understood.\cite{ParticleDataGroup:2022pth} 
Precise Belle II and STCF measurements of both branching ratios and spectral distributions 
will clarify whether there is indeed an anomaly,
making possible an accurate and reliable determination of $V_{us}$ from $\tau$ decays.

\subsection{Searching for hints of New Physics}

Precise measurements of hadronic $\tau$ decays can also be used as a probe of BSM physics. The sensitivity of the current $\tau$ data to non-standard interactions has been recently studied with effective field theory tools.\cite{Cirigliano:2018dyk,Gonzalez-Solis:2020jlh,Cirigliano:2021yto}
The leading-order charged-current interactions between quarks and leptons can be described through the effective four-fermion interaction\cite{Cirigliano:2021yto}
\begin{align}\label{eq:HeffSL}
\mathcal{L}_{\mathrm{eff}} \, & = \, 
-\frac{4 G_F}{\sqrt{2}}\,  V_{uD}\, \biggl\{
\left(\bar\ell_L\gamma_\mu\nu_\ell\right) \left[\left( 1 + \epsilon_L^{D\ell}\right) \left(\bar u_L\gamma^\mu D_L\right) + \epsilon_R^{D\ell}\left(\bar u_R\gamma^\mu D_R\right)\right]
\nonumber\\ &
\, + \frac{1}{2}\, 
\left(\bar\ell_R\nu_\ell\right)\;
\bar u\left( \epsilon_S^{D\ell}-\epsilon_P^{D\ell}\gamma_5\right) D
+\frac{1}{4}\, \hat\epsilon_T^{D\ell} \left(\bar\ell_R\sigma_{\mu\nu}\nu_\ell\right)
\left(\bar u_R\sigma^{\mu\nu} D_L\right)
\biggr\} + \mathrm{h.c.}\, ,
\end{align}
where $D=d,s$ denotes the down-type quark flavour, $\ell = e,\mu,\tau$ and $G_F\equiv G_{e\mu}$ is the Fermi constant measured in $\mu$ decay. Assuming CP conservation, new-physics contributions
are parametrized by real coefficients $\epsilon_X^{D\ell}$.

From a very comprehensive analysis of hadronic $\tau$ data, including inclusive and exclusive observables and strange and non-strange decay channels,  Ref.~\citen{Cirigliano:2021yto} has extracted the current constraints on the effective $\epsilon_X^{D\tau}$ couplings, which have been moreover combined with nuclear beta, baryon, pion and kaon decay data to obtain limits involving the three lepton families. This global fit reflects the current tension originating from the Cabibbo anomaly, indicating possible hints for BSM physics that need to be clarified. It also shows the high constraining power of the $\tau$ data and the very significant improvements that could be achieved in the future with high-precision experimental measurements.

\begin{figure}[tbh]
\centerline{\includegraphics[width=7cm]{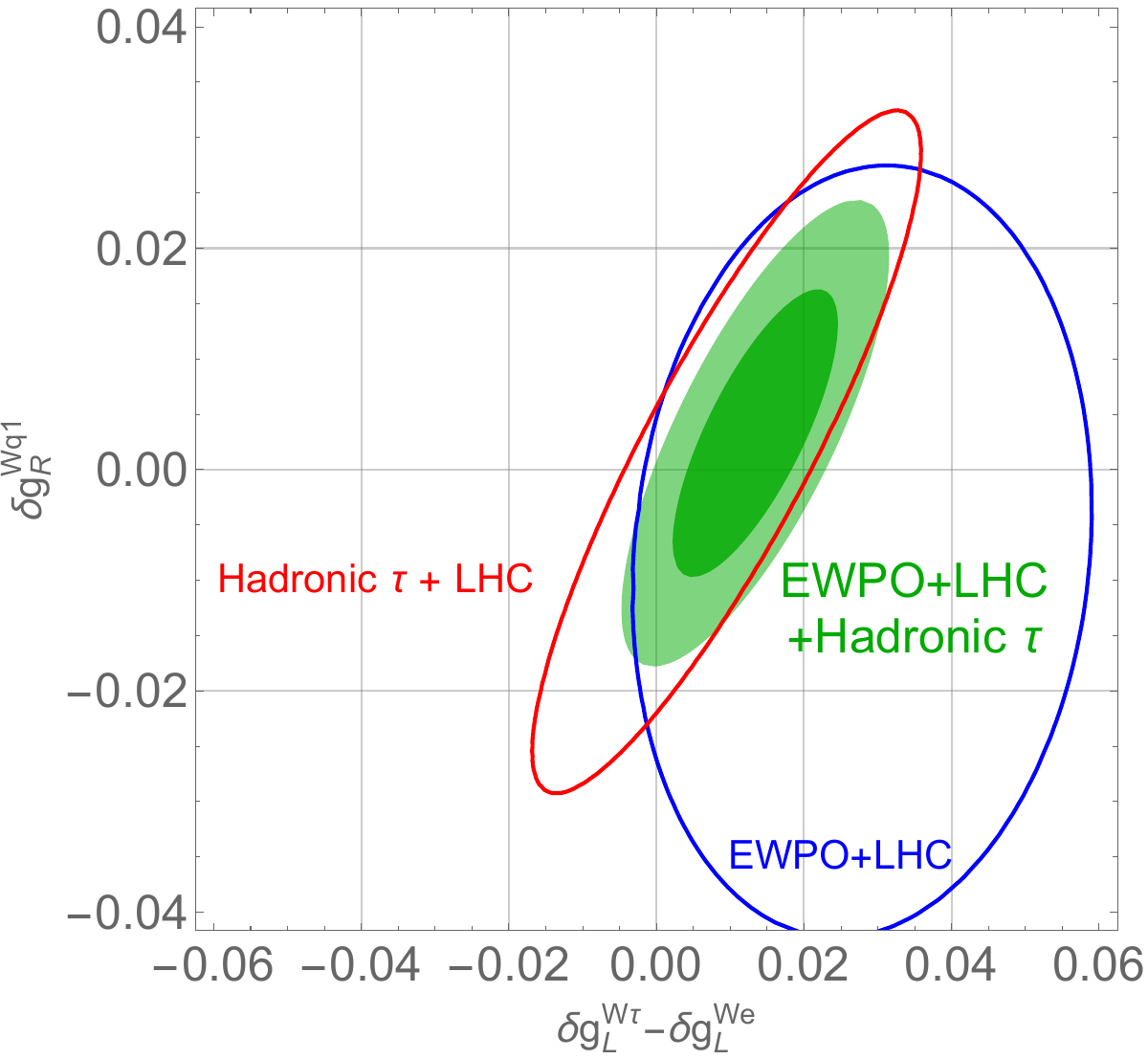}}
\caption{68\% and 95\% CL bounds on the $W$ vertex corrections, after using LHC input to constrain the SMEFT coefficients $[c_{\ell q}^{(3)}]_{\tau\tau 11}$ and $[c_{\ell q}^{(3)}]_{ee 11}$.\cite{Cirigliano:2018dyk} \label{fig:EWPO-LHC-Tau}}
\end{figure}

The effective couplings in Eq.~(\ref{eq:HeffSL}) can be easily matched to the SMEFT Wilson coefficients employed in high-energy effective field theory studies, making possible to combine in a model-independent way the constraints from low-energy experiments, electroweak precision observables (EWPO) and LHC searches.
The high-energy tail of $\tau\nu$ production at the LHC puts strong limits on some 4-fermion SMEFT couplings that also affect $\tau$ decays; once these LHC limits are imposed, the corrections to the SM $Wff'$ vertex are directly related to the $\epsilon_X^{D\tau}$ parameters:
$\delta g_R^{WD} = \epsilon_R^{D\tau}$, $\delta g_L^{W\tau}- \delta g_L^{We}\approx \epsilon_L^{D\tau}-\epsilon_L^{De}$. Fig.~\ref{fig:EWPO-LHC-Tau} compares the constraints on the $W$ vertex corrections obtained from hadronic $\tau$ decays (red) with those emerging from EWPO (blue), plus LHC data in both cases, exhibiting the high complementarity between the low-energy constraints and the information on BSM physics obtained at high-energy colliders.\cite{Cirigliano:2018dyk}
Combining all the information, one gets the much tighter limits shown in green.

\section{CP Violation}

CP violation is expected to be a sensitive probe for new physics because its fundamental origin remains still unknown. The SM mechanism of CP violation, based on the unique phase of the CKM  quark-mixing matrix, although successfully tested in many flavour experiments,  is unable to explain the matter-antimatter asymmetry of our universe.
In the SM with massless neutrinos, violations of the CP symmetry are predicted to be absent in the lepton sector, which makes $\tau$ decays a good laboratory to search for CP-violating signals from new dynamical sources.
A variety of CP-odd observables (rate, angular and polarization asymmetries,
triple products, Dalitz distributions, $\tau^+\tau^-$ correlations, etc.) can be exploited for this purpose.\cite{Pich:2013lsa}

The well-known (indirect) CP-violating contribution to $K^0$-$\bar K^0$ mixing should be reflected in a rate asymmetry between the $\tau^+\to\pi^+ K_S\bar\nu_\tau$ and 
$\tau^-\to\pi^- K_S \nu_\tau$ decays, which is precisely predicted by the CKM mechanism of CP violation:\cite{Bigi:2005ts,Grossman:2011zk}  $A_\tau^{\mathrm{SM}} = (3.6\pm 0.1)\cdot 10^{-3}$. Surprisingly, the BaBar collaboration measured in 2011 an asymmetry of the correct size but with the opposite sign:\cite{BaBar:2011pij}
\begin{equation}
A_\tau\,\equiv\,\frac{\Gamma(\tau^+\to\pi^+ K_S\bar\nu_\tau)-\Gamma(\tau^-\to\pi^- K_S \nu_\tau)}{\Gamma(\tau^+\to\pi^+ K_S\bar\nu_\tau)+\Gamma(\tau^-\to\pi^- K_S \nu_\tau)}
\, =\, (-3.6\pm 2.3\pm 1.1)\cdot 10^{-3}\, .
\end{equation}
This amounts to a $2.8\sigma$ discrepancy with the SM prediction that is quite difficult to understand. Using a model-independent effective-field-theory approach, it has been pointed out that this BaBar signal is in conflict with the constraints from the electric dipole moment (EDM) of the neutron and from $D^0$-$\bar D^0$ mixing, making highly improbable a BSM explanation from physics above the electroweak scale.\cite{Cirigliano:2017tqn,Rendon:2019awg,Chen:2019vbr}
Therefore, a confirmation of the signal would suggest the presence of BSM physics at very low scales so that the SMEFT bounds could be evaded.

The Belle collaboration has analysed a CP asymmetry constructed from the mean values of the $\cos{\beta}\cos{\psi}$ distribution in the $\tau^-\to\pi^- K_S \nu_\tau$ and $\tau^+\to\pi^+ K_S\bar\nu_\tau$ decays, 
where $\beta$ ($\psi$) is the angle between the $K_S$ ($\tau$) direction and the direction of the $e^+e^-$ CM frame, as observed from the hadronic rest system. Belle measured this angular asymmetry in four different bins of the $K_S\pi^\pm$ invariant-mass, finding that it is compatible with zero at the $10^{-2}$ level.\cite{Belle:2011sna} This angular asymmetry is sensitive to BSM scalar and tensor interactions.\cite{Chen:2021udz,Chen:2020uxi}
At the STCF, a sensitivity at the level of the SM prediction could be reached in both the rate and angular asymmetries.\cite{Sang:2020ksa}

The EDM of the $\tau$ is another (T-odd) observable of high interest, since it is expected to be more sensitive to BSM physics than the electron and muon EDMs, because of the heavier $\tau$ mass. Unfortunately, owing to the much more difficult experimental conditions, the current bounds on $d^\gamma_\tau$ from $e^+e^-\to\tau^+\tau^-$ are only at the level of  $10^{-17}\; e\,\mathrm{cm}$,\cite{Belle:2021ybo} compared to $10^{-29}$ ($10^{-19}$) for the electron (muon).\cite{ParticleDataGroup:2022pth} These limits could be improved by 2 or 3 orders of magnitude at the STCF.\cite{Achasov:2023gey,Wu:STCF}\footnote{The $e^+e^-\to\tau^+\tau^-$ data provides also information on the CP-conserving $\tau$ anomalous magnetic moment. A sensitivity on $|a_\tau^{\mathrm{NP}}|$ at the $10^{-5}$ level could be achieved at the STCF.\cite{Wu:STCF}}

With polarized  $e^+$ and $e^-$ beams one could build additional T-odd observables to search for CP-violating signals.\cite{Tsai:1994rc} In particular, a CP-odd asymmetry associated with the normal (to the scattering plane) polarization of a single $\tau$ turns out to be proportional to $\mathrm{Re} (d^\gamma_\tau)$.\cite{Bernabeu:2006wf} The STCF sensitivity could then reach $10^{-20}\; e\,\mathrm{cm}$.\cite{Achasov:2023gey}


\section{Lepton Flavour and Lepton Number Violation}

In the SM, the conservation of the total lepton $L=L_e+L_\mu+L_\tau$ and baryon numbers is an accidental symmetry of the dimension-4 Lagrangian. There exists a tower of $SU(2)_L\otimes U(1)_L$ gauge-invariant operators of higher dimensions, containing only SM fields, that break one or both of these symmetries; their contributions are suppressed by corresponding powers of the underlying new-physics scale.
%
The smallness of neutrino masses implies also a very strong suppression of any 
neutrinoless transition from one charged lepton flavour to another, leading to un-observably small rates. This suppression can be easily avoided in BSM models with sources of lepton flavour violation not related to $m_{\nu_i}$. 

The search for $\tau$ decays violating the lepton flavour or the total lepton number has then a superb potential to probe BSM physics. The high statistics that will be delivered by Belle II and the STCF could allow for a significant improvement of the current upper bounds on these decays,\cite{Achasov:2023gey} pushing them to the few$\times 10^{-10}$ level and providing the potential to probe new physics at scales much higher than those accessible through direct searches at ongoing and planned high-energy colliders.

\section{Summary}

A broad and very rich program of $\tau$ physics studies will be accessible at the future STCF. This superb facility will combine huge data samples with the particular advantages of the $\tau^+\tau^-$ threshold region, allowing for a tight control of backgrounds and systematic errors. A breakthrough in precision $\tau$ measurements could be realised, providing very relevant tests of the SM and constraining (or giving hints of) new BSM dynamics.

\section*{Acknowledgments}

Supported by MCIN/AEI/10.13039/501100011033, grant PID2020-114473GB-I00, and by Generalitat Valenciana, grant PROMETEO/2021 /071.

\bibliographystyle{ws-ijmpa}
\bibliography{TauReferences}
\end{document}